\documentclass[prd,onecolumn,superscriptaddress,nofootinbib,11pt]{revtex4-2}
\usepackage{url}

\usepackage{float}
\usepackage{epsfig}
\usepackage{amsfonts}
\usepackage{graphicx}
\usepackage{subfigure}
\usepackage{amsmath}
\usepackage{amssymb}
\usepackage{color}
\usepackage{bbm}
\usepackage{dcolumn}
\usepackage{bm}
\usepackage{ulem}
\usepackage{mathrsfs}
\usepackage{bbold}
\usepackage{datetime}
\usepackage{nameref}

\usepackage{rotating}
\usepackage[usenames,dvipsnames]{xcolor}
\usepackage[colorlinks=true,citecolor=Red,linkcolor=Green,urlcolor=Green]{hyperref}
\usepackage{lipsum} 
\usepackage{tikz,tikz-3dplot}
\usepackage{etoolbox} 
\usepackage[capitalize]{cleveref}
\usepackage{extarrows}
\usepackage{diagbox}

\def\bc{\begin{center}}

\def\ec{\end{center}}
\def\be{\begin{eqnarray}}
\def\ee{\end{eqnarray}}
\definecolor{dyellow}{rgb}{1.,0.8,.0}
\definecolor{myblue}{rgb}{.1,.1,.7}
\definecolor{dcyan}{rgb}{.0,.6,.6}
\definecolor{dmagenta}{rgb}{0.6,0.0,0.6}
\definecolor{brown}{rgb}{0.6,0.2,0.}
\definecolor{darkblue}{rgb}{.0,.0,0.5}
\definecolor{darkred}{rgb}{0.75,0.0,0.0}
\definecolor{orange}{rgb}{1.,.6,.0}
\definecolor{dorange}{rgb}{0.8,.4,.0}
\definecolor{darkgreen}{rgb}{0.0,0.6,0.0}
\definecolor{purple}{rgb}{.4,.0,.4}
\definecolor{lightgrey}{rgb}{0.7, 0.7, 0.7}
\definecolor{grey}{rgb}{0.4, 0.4, 0.4}

\usepackage{geometry}
\usepackage[font=small,labelfont=bf,justification=raggedright]{caption}

\newcommand{\xdownarrow}[1]{%
  {\left\downarrow\vbox to #1{}\right.\kern-\nulldelimiterspace}
}
\newcommand{\xuparrow}[1]{%
  {\left\uparrow\vbox to #1{}\right.\kern-\nulldelimiterspace}
}

\definecolor{myred}{RGB}{189, 38, 49}

\begin{document}
\title{Learning the Renyi entropy of multiple disjoint intervals in transverse-field quantum Ising models with restricted Boltzmann machine}
\author{Han-Qing Shi} \email{by2030104@buaa.edu.cn}
\affiliation{School of Physics and Optoelectronic Engineering, Beijing University of Technology, Beijing 100124, China}
\affiliation{Center for Gravitational Physics, Department of Space Science, Beihang University, Beijing 100191, China}
\author{Hai-Qing Zhang} \email{hqzhang@buaa.edu.cn}
\affiliation{Center for Gravitational Physics, Department of Space Science, Beihang University, Beijing 100191, China}
\affiliation{Peng Huanwu Collaborative Center for Research and Education, Beihang University, Beijing 100191, China}

\begin{abstract}
Renyi entropy with multiple disjoint intervals are computed from the improved swapping operations by two methods: one is from the direct diagonalization of the Hamiltonian and the other one is from the state-of-the-art machine learning method with neural networks. We use the paradigmatic transverse-field Ising model in one-dimension to demonstrate the strategy of the improved swapping operation. In particular, we study the second Renyi entropy with two, three and four disjoint intervals. We find that the results from the above two methods match each other very well within errors, which indicates that the machine learning method is applicable for calculating the Renyi entropy with multiple disjoint intervals. Moreover, as the magnetic field increases, the Renyi entropy grows as well until the system arrives at the critical point of the phase transition. However, as the magnetic field exceeds the critical value, the Renyi entropy will decrease since the system enters the paramagnetic phase. Overall, these results match the theoretical predictions very well and demonstrate the high accuracy of the machine learning methods with neural networks. 
\end{abstract}

\maketitle

\section{Introduction}
Quantum entanglement may be the most mysterious phenomenon in the nature \cite{nielsen2010quantum}. Entanglement entropy serves as a fundamental measure to quantify the quantum entanglement in many-body systems, characterized by the von Neumann entropy $S_A=-\text{Tr}(\rho_A\log\rho_A)$ in which $\rho_A$ is the reduced density matrix. \cite{horodecki2009quantum}. Among of these, Renyi entropy \cite{renyi1961measures}, which characterizes more information than the usual entropy, is an extension of the entanglement entropy. It provides a broader framework for measuring the uncertainty or randomness of a probability distribution and has found tremendous applications in various fields, including information theory, statistical physics, cryptography and etc \cite{beadle2008overview,fuentes2022renyi,skorski2015shannon}.

For a bipartite quantum system, we can separate its Hilbert space as $\mathcal{H}=\mathcal{H}_A\otimes\mathcal{H}_B$ where A and B denote the two spatially disjointed subsystems. Renyi entropy serves as a valuable tool for measuring the entanglement between the subsystem $A$ and its complement. For one-dimensional critical systems, the (Renyi) entanglement entropy has been analytically studied from the conformal field theory \cite{calabrese2009entanglement,calabrese2009entanglement2,calabrese2011entanglement,fagotti2010entanglement,alba2010entanglement}. However, until now there is still no analytical formula of the Renyi entropy for the non-critical systems, time evolution or non-equilibrium systems. In this case, we can resort to numerical simulations. For instance, high-precision approximated states can be obtained through quantum Monte Carlo simulations \cite{ceperley1986quantum,troyer2005computational}, machine learning methods \cite{carleo2019machine,Shi:2023bxx}, tensor networks \cite{orus2019tensor} and etc. Recently, Hastings et.al. \cite{hastings2010measuring} have introduced an approach to compute the Renyi entropy with one disjoint interval by evaluating the expectation value of a ``swapping operator". However, in the physical systems, such as quantum many-body systems \cite{zagoskin1998quantum}, holography \cite{nishioka2009holographic,faulkner2013entanglement,headrick2010entanglement} and etc, one may need to consider the Renyi entropy with multiple disjoint intervals, rather than considering a single interval. Under these circumstances, it is hard to calculate the Renyi entropy from the method introduced in \cite{hastings2010measuring}. Therefore, it is indispensable to develop a new method to calculate the Renyi entropy with multiple disjoint intervals. In this work, we develop a universal approach to study the generic $m$-th order Renyi entropy with multiple disjoint intervals, inspired from the similarities between the ``replica trick" in quantum field theory \cite{calabrese2009entanglement} and the swapping operation in quantum information theory. We dub this approach as ``improved swapping operation".

As a demonstration, we will investigate the Renyi entropy with multiple disjoint intervals in the one-dimensional transverse-field quantum Ising model (TFQIM) \cite{sachdev1999quantum} by using the improved swapping operations. In order to use the improved swapping operations, we need to first obtain the quantum states of the system, including either exact states or numerically approximated states. We will use two different methods to prepare the ground states of the system with various magnetic fields. One method is to directly diagonalize the Hamiltonian with the singular value decomposition (SVD) method \cite{greub2012linear}; The other one is to use the state-of-the-art machine learning methods with the neural networks \cite{carleo2019machine,carleo2017solving,Shi:2022dkw}. Nowadays, neural networks have been adopted in many areas, such as in the image and speech recognitions \cite{lecun2015deep}, the board game Go \cite{silver2016mastering}, the large language models \cite{chatGPT,deepseek} and etc. Among these, it also has been adopted in various fields of physics, such as nonlinear partial differential equations \cite{raissi2019physics}, quantum many-body physics \cite{carrasquilla2021neural,park2020geometry}, holographic entanglement entropy \cite{lam2021machine} and so on. Interested readers can refer to \cite{carleo2019machine,roberts2022principles}. In particular, neural networks can represent wave functions efficiently  and reduce the computational complexity dramatically \cite{le2008representational}. Therefore, quantum states in many-body physics can be represented by the neural networks, forming the so called ``neural quantum states" (NQS) \cite{carleo2019machine}. For instance, neural networks have been used to identify the phases and phase transitions in a variety of Hamiltonians \cite{carrasquilla2017machine}, and the phase transitions in the TFQIM were also studied through the NQS in \cite{shi2019neural}.

In this paper, we use the above two methods to study the second order Renyi entropy of two, three and four disjoint intervals in the ground states of the system by varying the magnetic fields. For simplicity, we work with the periodic boundary conditions, therefore, there will be a parity symmetry of the Renyi entropy along the size of the separations. Consequently, we find that the results obtained from the above two methods match each other very well within errors, which demonstrates that the machine learning method with neural networks is applicable for computing the Renyi entropy with multiple disjoint intervals. As the magnetic field approaches zero, the Renyi entropy is roughly a constant $\ln2$ which is consistent with the argument that in the thermodynamic limit, the ground state mainly contains two degenerated states: the spins all point up or down. As the magnetic field grows, the Renyi entropy will grow as well until it reaches the critical point which is in conformal field theory.  As the magnetic field continues increasing, the system will go beyond the critical regime, and the Renyi entropy will decrease since the system now enters the paramagnetic phase. As expected, the Renyi entropy will vanish if the magnetic field goes to infinity, since there will be only one configuration in which all spins will point to only one direction. 

This paper is arranged as follows: In Section \ref{swap} we introduce our improved swapping operation to compute the Renyi entropy with multiple disjoint intervals; In Section \ref{ML} we will concisely introduce the machine learning methods with the neural networks; Then we use two different methods to study the Renyi entropy with multiple disjoint intervals in TFQIM and compare them in Section \ref{application}; Finally, we draw our conclusions and discussions in Section \ref{conclusion}. In Appendix \ref{appendixa} we will show the detailed proof of the $m$-th order Renyi entropy with $n$ multiple disjoint intervals from the improved swapping operation; In Appendix \ref{appendixb}, we will show the details of the machine learning methods with neural networks to minimize the energy and briefly introduce the Metropolis Hastings algorithm.

\section{Swapping operation to compute the Renyi entropy with $n$ multiple disjoint intervals}
\label{swap}
The $m$-th order Renyi entropy is defined as:
\begin{equation}\label{smdef}
	S_m=\frac{1}{1-m}\ln \left({\rm Tr}(\rho_A^m)\right),
\end{equation}	
where the reduced density matrix $\rho_A$ is obtained by tracing the whole density matrix $\rho$ over the complement part of $A$, and $\rho^m_A$ represents the $m$-th power of $\rho_A$. It is known that in the limit of $m\to1$, the Renyi entropy will become the von Neumann entropy $S_1$ \cite{nielsen2010quantum}.  In a quantum system, due to the large dimension of the Hilbert space, directly computing the $m$-th power of the reduced density matrix $\rho_A^m$ is a formidable task. However, inspired from the swapping operation to compute the Renyi entropy with one disjoint interval in \cite{hastings2010measuring}, we can extend this approach to general $m$-th order Renyi entropy with generic $n$ disjoint intervals. This method will transform the computation of Renyi entropy to the evaluation of the expectation value of the swapping operator, which greatly simplifies the computation of Renyi entropy. We will concisely introduce the computation of $m$-th order Renyi entropy with $n$ disjoint intervals as follows (the details can be found in the Appendix \ref{appendixa}).

\begin{figure}[t]
	\centering
	\includegraphics[trim=5.cm 3.1cm 2.2cm 3.9cm, clip=true, scale=0.8]{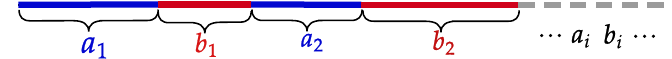}
	\caption{Schematic picture to illustrate the alternate arrangements of the intervals. The subsystem $A$ consists of the blue disjoint intervals $a_i$, i.e., $A=\cup_ia_i$ in which $(i=1,2,\cdots n)$ while the complement part $B=\bar A$ contains the red parts, i.e., $B=\cup_ib_i$. }
	\label{disjoint}
\end{figure}

In Fig.\ref{disjoint}, we assume the state of the entire system can be written as $|\Psi\rangle$. The subsystem $A$ (in blue) composes $A=\cup_i a_i$ while the complement part (in red) is $B=\cup_i b_i$ where $(i=1, 2,\cdots n)$. Their arrangements are alternately placed as $\{a_1b_1a_2b_2\cdots a_nb_n\}$. With this partition, the state of the whole system can be described as
\begin{eqnarray}
	|\Psi\rangle=\sum_{\textbf{ab}}C_{a_1b_1\cdots a_nb_n}\otimes_{i=1}^n(|a_i\rangle|b_i\rangle),
\end{eqnarray}	
in which $\sum_{\textbf{ab}}$ represents $\sum_{{a_1\cdots a_n;b_1\cdots b_n}}$ while $C_{a_1b_1\cdots a_nb_n}$ are the coefficients of the linear superposition.  $|a_i\rangle|b_i\rangle$ indicate the tensor product of the basis states of the $i$-th subsystem of $A$ and $B$ respectively, while $\otimes_{i=1}^n$ represents $n$ times of the tensor products of the states $|a_i\rangle|b_i\rangle$. 

\begin{figure}[h]
	\centering
	\includegraphics[trim=-2cm 0cm 0.cm 1.cm, clip=true, scale=0.8]{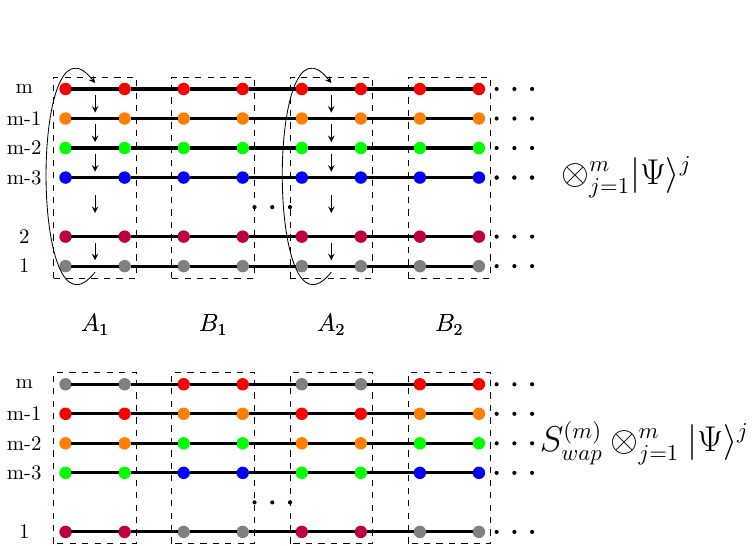}
	\caption{Schematic figure of the swapping operation on the $m$ replica states $\otimes_{j=1}^m|\Psi^j\rangle$. In the upper plot, we have separated the whole states into multiple parts, i.e., $A_1B_1A_2B_2\cdots$ which are enclosed in the dashed rectangles. The arrows appear only in the $A$'s parts, indicating that the swapping operation will exchange the $(j+1)$-st states to the $j$-th states, and the first states will be moved to the $m$-th states. The states in the $B$'s parts stay unchanged. After the swapping operation, the final states turn into $S_{\rm wap}^{(m)}\otimes_{j=1}^m|\Psi^{j}\rangle$ as the lower plot shows. }
	\label{swap_m}
\end{figure}

 In order to compute the value of $\text{Tr}(\rho_A^m)$, we need to copy the state $|\Psi\rangle$ for $m$ times and then tensor product them together, i.e., $\otimes_{j=1}^m|\Psi^j\rangle$ in which the index $j$ indicates the $j$-th replica of the state $|\Psi\rangle$.\footnote{It should be stressed that $m$ replicas of the state $|\Psi\rangle$ resemble the replica trick in the path integral in quantum field theory. There, people also copy the fields $m$ times as $m$ sheets and then connect them by some particular boundary conditions to compute the Renyi entropy \cite{calabrese2009entanglement}. }  The $m$-th order swapping operator ${S}_{\rm wap}^{(m)}$ will act on the $m$ copies of the states $\otimes_{j=1}^m|\Psi^j\rangle$ by replacing the state $|a_i^{j+1}\rangle$ in the $(j+1)$-st replica subsystem $A^{j+1}$ with the state $|a_i^{j}\rangle$ in the $j$-th replica subsystem $A^j$, i.e., $|a_i^{j+1}\rangle\to |a_i^{j}\rangle$. It means this kind of swapping operation is done for every two adjacent replicas. However, it should be noted that the state in the first subsystem $A^1$ must be replaced with the state in the $m$-th (the last) subsystem $A^m$, i.e., $|a_k^1\rangle\to|a_k^m\rangle$. Since we are considering the Renyi entropy of the subsystem $A$, the swapping operator ${S}^{(m)}_{\rm wap}$ will not act on the states $|b_i^j\rangle$ in the complement subsystem $B$. Please refer to the Fig.\ref{swap_m} for the illustrative explanations of the swapping operation.  Then, the expectation value of the swapping operator $S_{\rm wap}^{(m)}$ under the $m$ copied states can be straightforwardly computed $\langle S_{\rm wap}^{(m)}\rangle\equiv\langle\otimes_{j'=1}^m\Psi^{j'}|S_{\rm wap}^{(m)}|\otimes_{j=1}^m\Psi^{j}\rangle$. 

On the other hand, we can evaluate $\text{Tr}(\rho_A^m)$ by tracing the reduced density matrix $\rho_A^m$ over the parts of $A$ system. As a result, we find the following equivalence
\begin{equation}
	\text{Tr}(\rho_A^m)=\langle S_{\rm wap}^{(m)}\rangle.
\end{equation}
Therefore, from the definition of the Renyi entropy \eqref{smdef} the computation of the $m$-th Renyi entropy $S_m$ with $n$ disjoint intervals is transformed to computing the expectation values of the swapping operator,  
\begin{equation}
	S_{m}=\frac{1}{1-m}\ln(\langle S_{\rm wap}^{(m)}\rangle).
\end{equation}

It should be stressed that the above method applies for various forms of the quantum states, such as the ground states, the excited states, the dynamical states, and etc. For simplicity, we can directly diagonalize the Hamiltonian to get the ground states of the system, then we can calculate the expectation value of the swapping operator $S_{\rm wap}^{(m)}$ to obtain the exact value of the $m$-th order Renyi entropy with $n$ multiple disjoint intervals. However, diagonalizing a Hamiltonian can only be applied for small systems. For larger systems, the dimension of the Hilbert space will be exponentially increased, which makes the diagonalization of the Hamiltonian very difficult. Fortunately, the neural network can be used to approximate the ground states of the Hamiltonian \cite{carleo2017solving,carleo2019machine}. At this point, one can use the quantum Monte Carlo method to sample the system's configurations of the states. As a result, even for larger systems we can still calculate the Renyi entropy of multiple disjoint intervals within the allowable errors, which greatly reduces the computational complexity of the Renyi entropy.

\section{Machine learning methods with neural networks}
\label{ML}
\begin{figure}[h]
	\centering
	\includegraphics[trim=0cm 0cm 0cm 0cm, clip=true, scale=0.7]{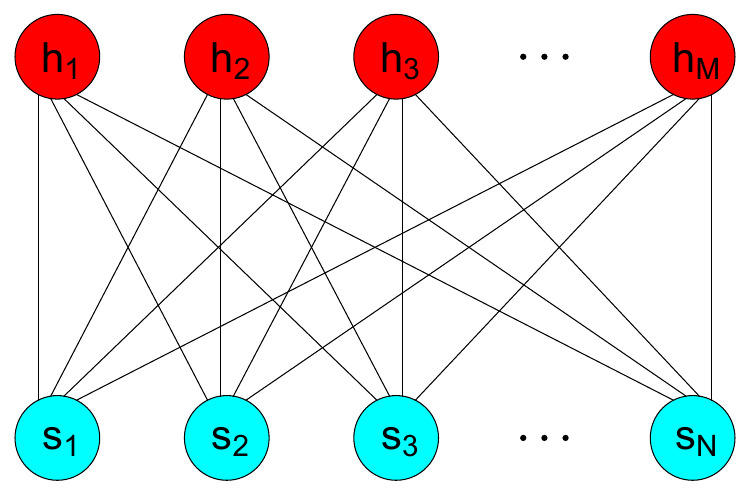}
	\caption{Schematic figure of the structure of the RBM.  The visible layer (in green) consists of $N$ visible neurons $s_j$ where $(j=1, 2, \cdots N)$, each of which corresponds to the spin direction $s_j=\{-1,1\}$ at every site of the system, while the hidden layer (in red) is composed of $M$ hidden neurons $h_i$ where $(i=1, 2, \cdots M)$, which take values as $h_i=\{-1,1\}$.}
	\label{RBM}
\end{figure}

In this section we will introduce how to use the machine learning methods with neural networks to prepare the ground states of the quantum systems, in order to compute the expectation value of the swapping operator $\langle S_{\rm wap}^{(m)}\rangle$. We call the quantum states prepared by the neural networks as neural-network quantum states (NQS) and written as  $\Psi_{\!N\!N}(s,\mathcal{W})$, where $s=(s_1,s_2\dots,s_N)$ denotes the quantum configurations at each site, and $\mathcal{W}$ is the neural network parameter. NQS is a powerful approximated wave function described by a neural network, where the parameter $\mathcal{W}$ controls the shape of the function of the system.  Therefore, the goal of machine learning with the neural network is to train the system and find a suitable parameter $\mathcal{W}$ to describe the state of the system. In our paper, we have chosen a structure that is highly compatible with quantum spin systems, i.e., the restricted Boltzmann machine (RBM) as the neural network for NQS \cite{carleo2017solving,fischer2012introduction}. \footnote{The RBM was initially proposed in \cite{smolensky1986information} and then promoted by Hinton in \cite{hinton2002training}. RBM has received many applications in classification \cite{larochelle2008classification}, feature learning \cite{coates2011analysis}, immunology \cite{bravi2023transfer}, quantum-many body systems and etc. \cite{carleo2017solving,fischer2012introduction}. } The RBM consists of two layers, one is the visible layer with $N$ visible neurons (denoting as $s_j$, $(j=1, 2, \cdots, N)$) and the other layer is called the hidden layer which contains $M$ hidden neurons (denoting as $h_i$, $(i=1, 2, \cdots, M)$), please refer to Fig.\ref{RBM}.  RBM have connections between the visible and hidden layers but do no have internal connections between neurons in each layer. We have chosen the number of hidden neurons to be four times than the visible neurons, i.e., $M=4N$, which is sufficient to represent the states of the system we are studying \cite{Shi:2022dkw,Shi:2023bxx}. According to the structure of the neural networks, the quantum state can be described as
\begin{eqnarray}\label{psinn}
\Psi_{\!N\!N}(s,\mathcal{W})\!=
\sum_{\{h\}}\exp\!\left[\sum_ja_js_j+\sum_ib_ih_i+\sum_{i,j}w_{ij}h_is_j\right],
\end{eqnarray}
where $s=\{s_j\}$ denotes the spin configurations (or the visible neurons in Fig.\ref{RBM}) and $\mathcal{W}=\{a_j,b_i,w_{ij}\}$ are the neural network parameters. We set the network parameters $\mathcal{W}$ to be complex so that $\Psi_{\!N\!N}$ can represent both the amplitudes and phases of the states. The value of $s_j$ and $h_i$ can be either $1$ or $-1$, corresponding respectively to the spin up or down. Since the hidden neurons are auxiliary structures and there are no interactions within the hidden layer, we can traced out the hidden variables with the identity $2\cosh(x)=e^x+e^{-x}$ in the first step. Thus, Eq.\eqref{psinn} becomes
\begin{equation}
\Psi_{\!N\!N}(s,\mathcal{W})=e^{\sum_ja_js_j}\prod_{i=1}^M2\cosh[b_i+\sum_jw_{ij}s_j].
\end{equation}

In the machine learning methods, our goal is to find the network parameters $\mathcal{W}$ that can represent the ground states of the system. To achieve this goal, we adopt the stochastic reconfigurations (SR) method to minimize the expectation value of the system's energy $\langle E\rangle=\langle\Psi_{\!N\!N}|H|\Psi_{\!N\!N}\rangle/\langle\Psi_{\!N\!N}|\Psi_{\!N\!N}\rangle$ (The details of the SR method are provided in the Appendix \ref{appendixb}). When training NQS, the SR method can provide a rapid path to the lowest energy state in the parameter space. However, along this path, it is necessary to estimate the system's energy expectation value at each step. In order to avoid the difficulties brought by the exponentially large Hilbert space, we need to use a sampling method --- the Metropolis Hastings algorithm to approximate the system's energy expectation value. The specific steps of this method are also provided in the Appendix \ref{appendixb}.
Additionally, we reduce the number of independent parameters from the symmetry of the system. This improvement not only increases computational efficiency but also enhances the stability of the program. The symmetry of the system refers to the invariance of the wave function under a translational transformation $\hat{T}$. Therefore, we can impose the constraint $\Psi_{\!N\!N}(s,\mathcal{W})=\Psi_{\!N\!N}(\hat{T}s,\mathcal{W})$ into the parameters $\mathcal{W}$ to reduce the number of independent parameters \cite{codesurl}. 

\section{Application in one-dimensional TFQIM}
\label{application}
In this section, we will compute the Renyi entropy with multiple disjoint intervals by the improved swapping operation.
For simplicity, we consider the paradigmatic one dimensional TFQIM with periodic boundary conditions. The Hamiltonian of TFQIM with $N$ sites is,
\begin{equation}\label{TFQIM}
	H=-J\sum_{i=1}^N(\sigma_i^z\sigma_{i+1}^z+h\sigma_i^x),
\end{equation}
where $\sigma_i^z$ and $\sigma_i^x$ are respectively the $z$ and $x$-components of the Pauli matrices at the $i$-th site, $J$ represents the coupling strengths between the nearest-neighboring sites (we have set $J=1$ in the computation), and $h$ denotes the strength of transverse magnetic field along $x$-direction. Because of the periodic boundary conditions, the $(N+1)$-st site is equivalent to the first site, viz., $\vec{\sigma}_{N+1}=\vec{\sigma}_1$.   

In order to get the expectation values of the swapping operator $\langle S_{\rm wap}^{(m)}\rangle$, we need to obtain the states of the system at first. For simplicity, we will work in the ground states of the system in different phases by varying the magnetic field strength $h$. There exists a phase transition from the ferromagnetic to paramagnetic while the critical value of $h$ is $h_c=1$. For these ground states, without the loss of generality, we will compute the second order Renyi entropy $S_2$ with two, three and four disjoint intervals, in particular from two different methods.\footnote{In principle, our method from the swapping operation can be used to compute general order of the Renyi entropy with general multiple disjoint intervals. However, due to the capacity and efficiency of the computer, we only study the simplest cases to compute $S_2$ with two, three and four disjoint intervals, which can still demonstrate the applicability of our methods in computing Renyi entropy.} The first method is to use the above machine learning methods with the neural network to prepare the approximate ground states of the systems. We choose the spin $z$-direction as the basis state to represent the system by using the RBM. Then, we can obtain the ground states by minimizing the expectation value of the energy with the SR method, please refer to Appendix B.  The second method is to directly diagonalize the Hamiltonian to get the states of the system: For a finite spin chain, the dimension of Hilbert space of the Hamiltonian grows exponentially with the size of the system. We can consider the following eigenvalue equation,
\begin{equation}
	H|\Psi_i\rangle=E_i|\Psi_i\rangle,
\end{equation}
in which $E_i$ and $|\Psi_i\rangle$ are respectively the $i$-th energy eigenvalue and eigenvector. In the TFQIM model \eqref{TFQIM}, we have adopted up to $N=24$ sites. The Hamiltonian matrices in TFQIM model are typically sparse matrices, hence, we can use the singular value decomposition (SVD) method to diagonalize the Hamiltonian matrix directly with the dimension up to $10^8\times 10^8$ (since $2^{24}\approx 1.6777\times10^{7}$). After many trials, we find that this capacity can be readily realized on a personal computer. After diagonalizing the Hamiltonian, we then can find the ground states of $H$ readily. For quantum systems of this scale, we have two distinct approaches to compute the expectation value of the swapping operator $\langle S_{\rm wap}^{(m)}\rangle$: The first strategy involves extracting all necessary parameters directly from the obtained ground state wavefunction to calculate the expectation value analytically. The alternative approach employs Monte-Carlo sampling techniques to statistically estimate the expectation value. In our current numerical implementation, we have adopted the first strategy at this system size. However, we note that the second Monte-Carlo based approach would be more appropriate and scalable for larger systems beyond the current scope.

\subsection{$S_2$ with two disjoint intervals}\label{sec2}

\begin{figure}[h]
	\centering
	\includegraphics[trim=14.2cm 27.1cm 15.9cm 29.2cm, clip=true, scale=0.2]{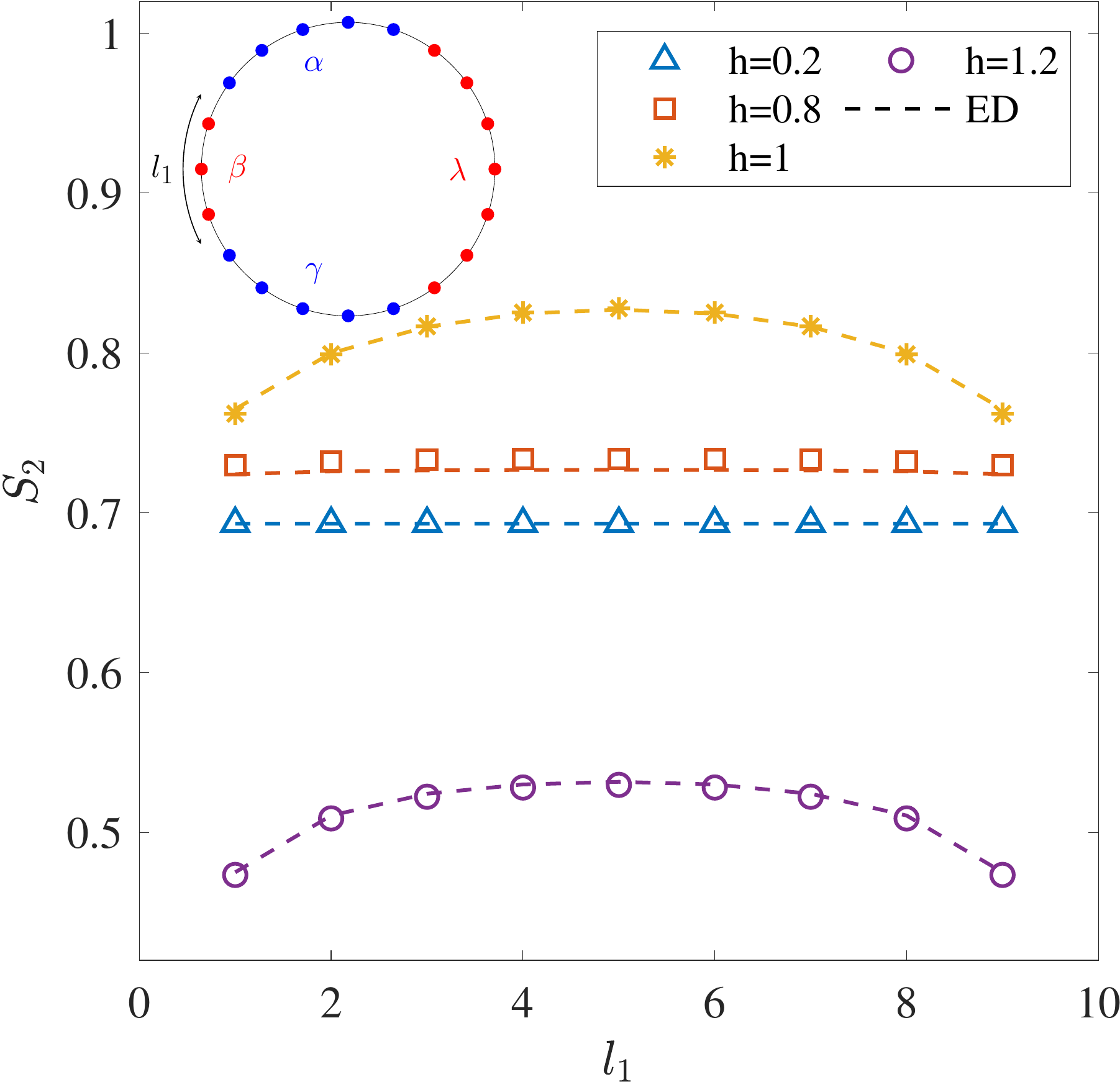}~~~~
	\includegraphics[trim=14.2cm 27.1cm 15.9cm 29.2cm, clip=true, scale=0.2]{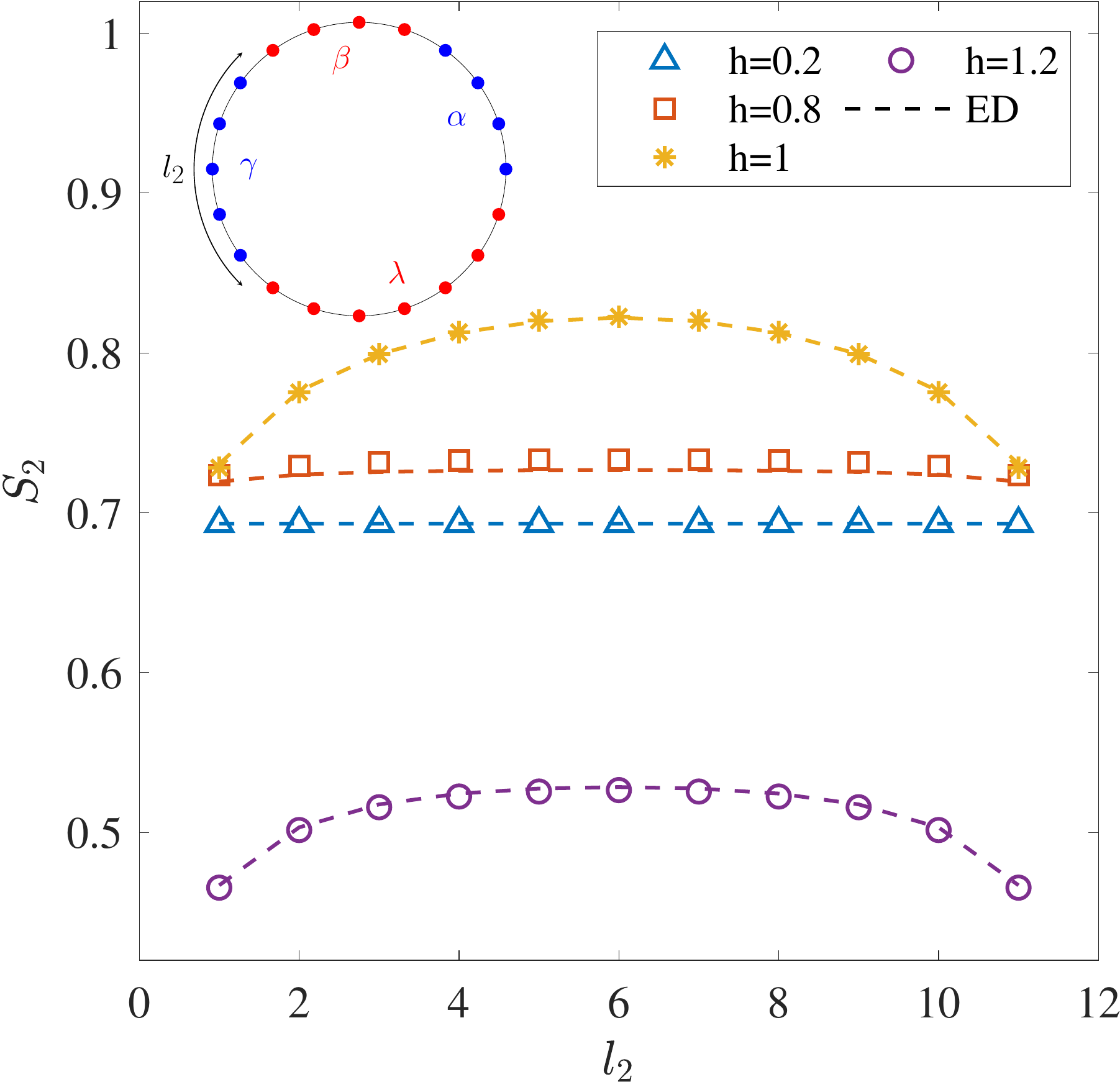}
	\caption{(Left) Renyi entropy $S_2$ with two disjoint intervals ($A=\alpha\cup\gamma$) against $l_1$, the length of $\beta$ interval; (Right) Renyi entropy $S_2$ with two disjoint intervals ($A=\alpha\cup\gamma$) against $l_2$, the length of $\gamma$ interval. In both panels, the circles in the top-left corners are the schematic pictures to show the arrangements of the disjoint intervals. Meanwhile, different styles of the points represent the Renyi entropy obtained from the machine learning methods with neural network for various magnetic fields, while the dashed lines correspond to the diagonalization of the Hamiltonian from the SVD method. They match each other very well. }
	\label{interval}
\end{figure}

First, we consider dividing the system into four subregions. Subsystem $A$ and its complement $B$ each contains two disjoint intervals, please refer to the top-left corner in the left plot of Fig.\ref{interval}. Subsystem $A$ (in blue) includes subregions $\alpha$ and $\gamma$, while the complement part $B$ (in red) includes subregions $\beta$ and $\lambda$. In this scenario, we choose the size of the whole system as 20 sites. We fix the size of each subregions $\alpha$ and $\gamma$ to be five, and vary the separation (the length of $\beta$ region which denotes as $l_1$) between these two subregions $\alpha$ and $\gamma$. The second order Renyi entropy $S_2$ will be computed for the subsystem $A$ by varying the the separation $l_1$ with various magnetic field strengths $h$. In the left panel of Fig.\ref{interval}, we show the second order Renyi entropy $S_2$ against the length $l_1$ under different transverse magnetic field strength $h$. Due to the periodic boundary conditions, it is obvious to see that changing the size of $\beta$ is equivalent to changing the size of $\lambda$. Therefore, there should be a parity symmetry $l_1\leftrightarrow (10-l_1)$ of the Renyi entropy. Hence, it greatly simplifies our computation due to this kind of parity symmetry. In plotting the left panel of Fig.\ref{interval}, we only plot the left part ($l_1\leq5$) of the picture, and then just copy the left part to the right part.\footnote{In the following figures, we also use this kind of strategy to plot them.} These lines should have a parity symmetry along the vertical line $l_1=5$.  

In the left panel of Fig.\ref{interval} the data points, i.e., yellow asterisks, red squares, blue triangles and purple circles are the numerical results from the machine learning methods with neural networks. The corresponding colorful dashed lines are from the methods of diagonalizing the Hamiltonian with the SVD.  We define the relative error of Renyi entropy as $\epsilon=|S_{\rm ML}-S_{\rm SVD}|/S_{\rm SVD}$, where $S_{\rm ML}$ is the Renyi entropy from the machine learning methods while $S_{\rm SVD}$ is the Renyi entropy from the diagonalization of the Hamiltonian by SVD method. The relative errors for these two methods are exhibited in Table \ref{tab1}.  We can see that these two methods match each other with high precision.  
\begin{table}[h]
	\centering
	\begin{tabular}{|c|c|c|c|c|c|}
		\hline
		\diagbox{$h$}{$l_1$}&$1$&$2$&$3$&$4$&$5$\\
		\hline
		$0.2$&$3.89\times10^{-6}$&$4.27\times10^{-6}$&$4.36\times10^{-6}$&$4.38\times10^{-6}$&$4.39\times10^{-6}$\\
		$0.8$&$7.51\times10^{-3}$&$8.54\times10^{-3}$&$9.05\times10^{-3}$&$9.31\times10^{-3}$&$9.39\times10^{-3}$\\
		$1.0$&$3.74\times10^{-3}$&$9.06\times10^{-4}$&$3.21\times10^{-4}$&$8.01\times10^{-4}$&$9.62\times10^{-4}$\\
		$1.2$&$3.52\times10^{-3}$&$3.52\times10^{-3}$&$3.56\times10^{-3}$&$3.58\times10^{-3}$&$3.57\times10^{-3}$\\
		\hline
	\end{tabular}
	\caption{The relative errors $\epsilon$ between the machine learning methods and the SVD methods for computing $S_2$ with two disjoint intervals.  The first row represents the separation size $l_1$ while the first column represents the magnetic field strength $h$. This table corresponds to the left plot of Fig.\ref{interval}.}\label{tab1}
\end{table}

From the left plot of Fig.\ref{interval}, we observe that as the magnetic field approaches zero ($h=0.2$), the entropy tends to a constant value of $\ln 2\approx0.69$. This occurs because the system approaches the ferromagnetic phase where, in the thermodynamic limit, it would exhibit two degenerate ground states corresponding to the two spin configurations  $\lvert\uparrow\uparrow\cdots\uparrow\rangle$ and $\lvert\downarrow\downarrow\cdots\downarrow\rangle$. However, we note that for our finite-size system ($N=24$) at $h=0.2$, the true ground state is actually unique due to quantum fluctuations, though the entropy value still reflects the near-degeneracy of the ground states. Then the Renyi entropy will increase with the magnetic field until it reaches its maximum at the critical magnetic field $h_c=1$. 
At the critical point, $S_2$ increases with $l_1$ until it reaches the symmetry axis $l_1=5$. As the magnetic field $h$ exceeds the critical value, for instance $h=1.2$, the entropy will decrease compare to those at the critical point, since now the system enters the paramagnetic phase. If $h\to\infty$, the Renyi entropy will vanish as expected, because there will be only one configuration that all the spins will point to the $x$-direction \cite{calabrese2009entanglement}.

In the right panel of Fig.\ref{interval}, we keep the size of the regions $\alpha$ and $\beta$ as four, respectively. Then we change the size of $\gamma$, which is denoted as $l_2$. Please be noted that in this case the subsystem $A$ is still composed of the blue subregions, i.e., $A=\alpha\cup\gamma$.  The relationship between Renyi entropy $S_2$ and the size of $\gamma$ is shown in the right plot of Fig.\ref{interval}. As before, due to the periodic boundary conditions, there is also a parity symmetry $l_2\leftrightarrow (12-l_2)$. 
In the right plot, the data points, i.e., yellow asterisks, red squares, blue triangles and purple circles are the numerical results from the machine learning methods with neural networks, while the corresponding colorful dashed lines are from the the SVD methods. The relative errors $\epsilon$ between the two methods are shown in the Table \ref{tab2}, from which we can see that these two methods align with each other very well.
\begin{table}[h]
	\centering
		\begin{tabular}{|c|c|c|c|c|c|c|}
		\hline
		\diagbox{$h$}{$l_2$}&$1$&$2$&$3$&$4$&$5$&$6$\\
		\hline
$0.2$&$3.83\times10^{-6}$&$4.23\times10^{-6}$&$4.33\times10^{-6}$&$4.37\times10^{-6}$&$4.38\times10^{-6}$&$4.38\times10^{-6}$\\
$0.8$&$5.71\times10^{-3}$&$7.36\times10^{-3}$&$8.28\times10^{-3}$&$8.84\times10^{-3}$&$9.14\times10^{-3}$&$9.24\times10^{-3}$\\
$1.0$&$3.24\times10^{-3}$&$9.36\times10^{-4}$&$2.37\times10^{-6}$&$4.47\times10^{-4}$&$6.37\times10^{-4}$&$6.86\times10^{-4}$\\
$1.2$&$3.18\times10^{-3}$&$3.31\times10^{-3}$&$3.44\times10^{-3}$&$3.53\times10^{-3}$&$3.57\times10^{-3}$&$3.59\times10^{-3}$\\
		\hline
	\end{tabular}
	\caption{The relative errors $\epsilon$ between the machine learning methods and the SVD methods for computing $S_2$ with two disjoint intervals.  The first row represents the size of $\gamma$, i.e. $l_2$, while the first column represents the magnetic field strength $h$. This table corresponds to the right plot of Fig.\ref{interval}.}\label{tab2}
\end{table}

As before, we see that when $h$ is small as $h=0.2$, the Renyi entropy is roughly a constant $\ln2$. When $h$ increases the entropy increases as well until it arrives at the maximum when the system is at the critical point of the phase transition. Then
we continue increasing the magnetic field, the system will go beyond the critical regime and enters the paramagnetic phase. In this case the Renyi entropy $S_2$ will decrease as the results for $h=1.2$ shows. As expected, when $h\to\infty$ the Renyi entropy will vanish since spins will all point to the $x$-direction.

\subsection{$S_2$ with three and four disjoint intervals}

\begin{figure}[h]
	\centering
	\includegraphics[trim=14.2cm 27.1cm 15.9cm 28.8cm, clip=true, scale=0.2]{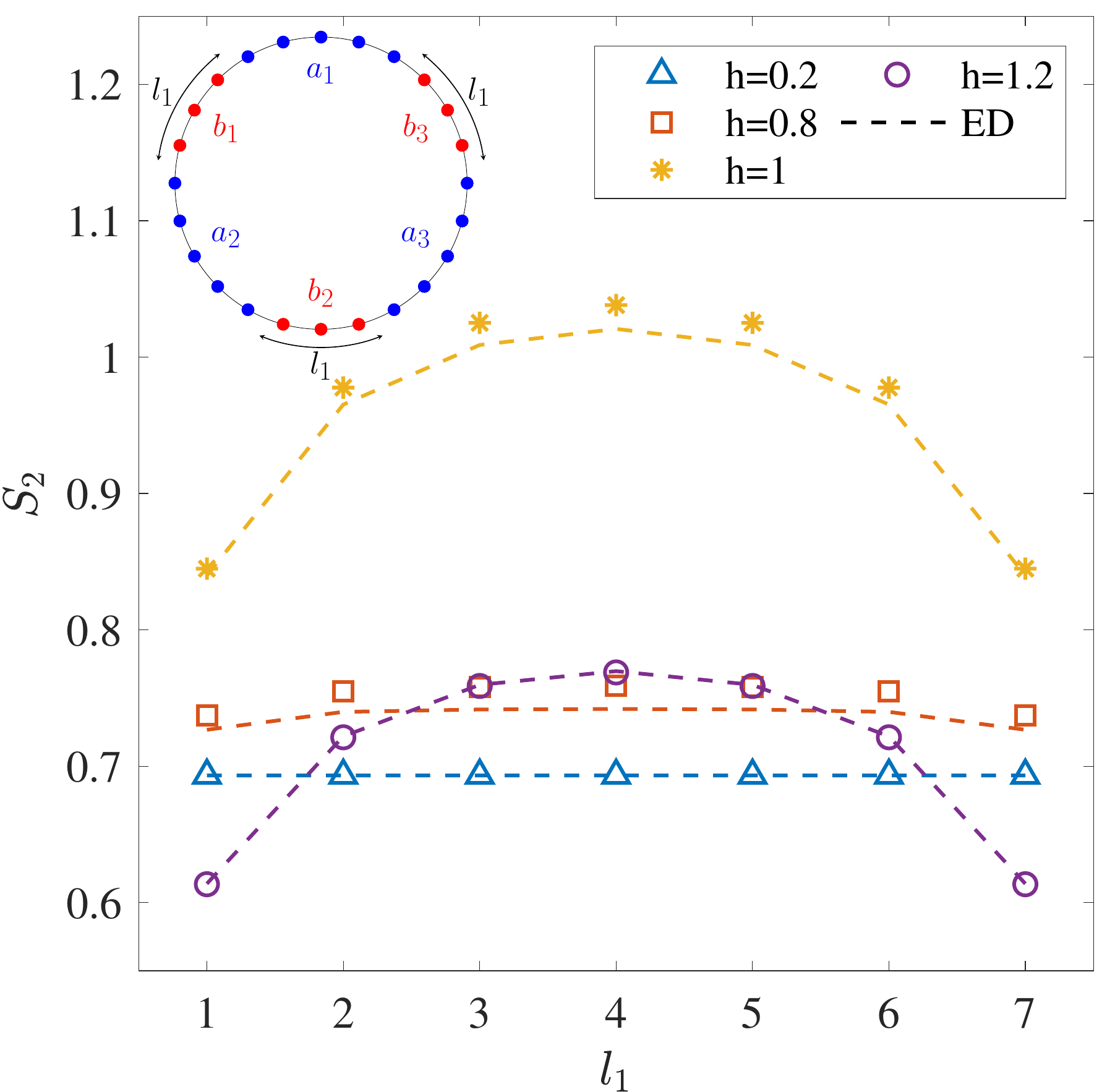}~~~~
	\includegraphics[trim=14.2cm 27.1cm 15.9cm 28.8cm, clip=true, scale=0.2]{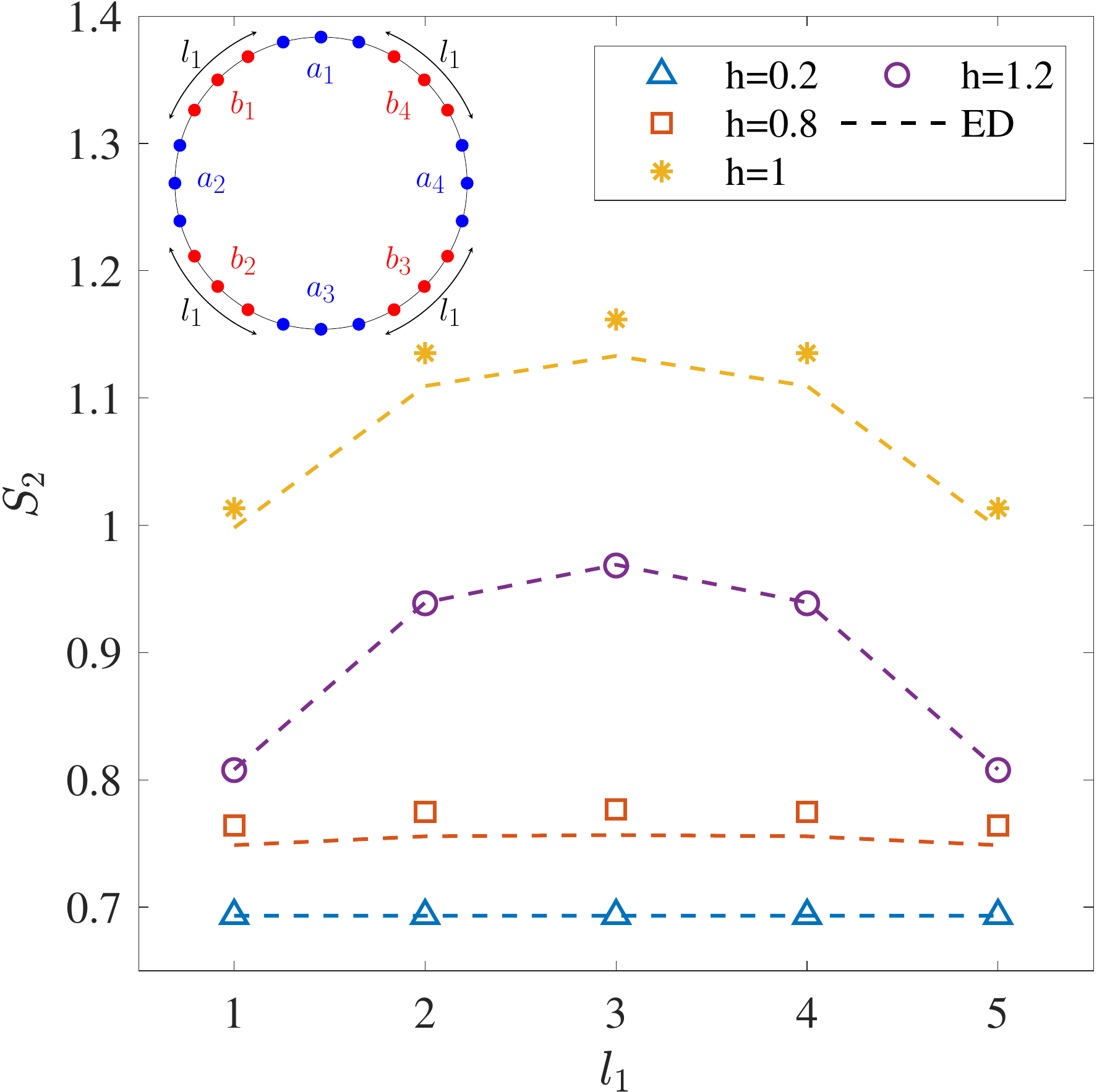}
	\caption{(Left) Renyi entropy $S_2$ with three disjoint intervals $A=a_1\cup a_2\cup a_3$ against $l_1$, the size of each $b_i$ $(i=1, 2, 3)$. (Right) Renyi entropy $S_2$ with four disjoint intervals $A=a_1\cup a_2\cup a_3\cup a_4$ against $l_1$, the size of each $b_i$ $(i=1, 2, 3, 4)$. In both panels, the circles in the top-left corners are the schematic pictures to show the arrangements of the disjoint intervals. Meanwhile, different styles of the points represent the Renyi entropy obtained from the machine learning methods with neural network for various magnetic fields, while the dashed lines correspond to the diagonalization of the Hamiltonian from the SVD method. They match each other very well within errors.}
	\label{b6}
\end{figure}

For completeness, we also consider dividing the system into more subregions. Now, we set the system with $24$ sites, and divide it into six parts and eight parts respectively, please refer to the circles in the top-left corner of the two panels in Fig.\ref{b6}. The subsystem $A$ contains the blue parts, i.e., $A=\cup_ia_i$, and the complement part $B$ contains the red parts, i.e., $B=\cup_ib_i$ where $i=1,2,3$ (for the left plot) or $i=1,2,3,4$ (for the right plot). For simplicity, we equally partition the subregions, which means that each $a_i$ part has the same size and each $b_i$ part has the same size as well. And we study the Renyi entropy $S_2$ by changing the size of $b_i$ (denoted as $l_1$) with various magnetic field strength $h$. Similar to Fig.\ref{interval}, all the different styles of the points represent the numerical data from the machine learning methods, while the dashed lines represent the method from diagonalization of the Hamiltonian by SVD. 

Obviously, from the left plot of Fig.\ref{b6}, there is a parity symmetry $l_1\leftrightarrow (8-l_1)$ due to the periodic boundary conditions and the equally partitioned subregions of $a_i$ and $b_i$. In the same manner, the right plot of Fig.\ref{b6} has a parity symmetry $l_1\leftrightarrow (6-l_1)$. As the magnetic field is close to zero such as $h=0.2$, the Renyi entropies $S_2$ are roughly constants $\approx\ln2$, which can be explained similarly as that in the preceding subsection. Then, as $h$ increases, the Renyi entropies grow as well until the system arrives at the critical point of the phase transition. As $h$ goes beyond the critical value $h_c=1$, the entropies decrease again since the system enters into the paramagnetic phase. As expected, the entropies will vanish as $h\to\infty$. \\

\begin{minipage}{\textwidth}
\!\!\!\!\!\!\!\!\!\!
\begin{minipage}[t]{0.6\textwidth}
\makeatletter\def\@captype{table}
	\begin{tabular}{|c|c|c|c|c|}
		\hline
		\diagbox{$h$}{$l_1$}&$1$&$2$&$3$&$4$\\
		\hline
		$0.2$&$2.71\times10^{-6}$&$4.02\times10^{-6}$&$4.34\times10^{-6}$&$4.41\times10^{-6}$\\
		$0.8$&$1.44\times10^{-2}$&$2.01\times10^{-2}$&$2.22\times10^{-2}$&$2.28\times10^{-2}$\\
		$1.0$&$5.99\times10^{-3}$&$1.29\times10^{-2}$&$1.61\times10^{-2}$&$1.70\times10^{-2}$\\
		$1.2$&$6.34\times10^{-4}$&$9.82\times10^{-4}$&$1.11\times10^{-3}$&$1.22\times10^{-3}$\\
		\hline
	\end{tabular}
	\caption{The relative errors $\epsilon$ between the machine \\ learning methods and the SVD methods for computing $S_2$ \\ with three disjoint intervals. The first row $l_1$ represents the \\ size of $b_i$ $(i=1, 2, 3)$, while the first column represents the \\ magnetic field strength $h$. This table corresponds to the left \\plot of Fig.\ref{b6}.}\label{tab3}
\end{minipage}\!\!\!\!\!\!\!\!
\begin{minipage}[t]{0.43\textwidth}
\makeatletter\def\@captype{table}
	\centering
	\begin{tabular}{|c|c|c|c|}
		\hline
		\diagbox{$h$}{$l_1$}&$1$&$2$&$3$\\
		\hline
		$0.2$&$5.70\times10^{-6}$&$7.42\times10^{-6}$&$7.73\times10^{-6}$\\
		$0.8$&$2.09\times10^{-2}$&$2.54\times10^{-2}$&$2.64\times10^{-2}$\\
		$1.0$&$1.53\times10^{-2}$&$2.32\times10^{-2}$&$2.53\times10^{-2}$\\
		$1.2$&$4.90\times10^{-4}$&$6.94\times10^{-4}$&$6.32\times10^{-4}$\\
		\hline
	\end{tabular}
	\caption{The relative errors $\epsilon$ between the machine learning and the SVD methods for $S_2$ with four disjoint intervals. The first row $l_1$ is the size of $b_i$ $(i=1, 2, 3, 4)$, and the first column is the magnetic field strength $h$. This table corresponds to the right plot of Fig.\ref{b6}.}\label{tab4}
\end{minipage}
\end{minipage}

From Fig.\ref{b6} we see that the data points and the dashed lines match each other very well although at some points there are a little bit discrepancies. We list the relative errors $\epsilon$ in the Table \ref{tab3} (corresponds to the left panel of Fig.\ref{b6}) and Table \ref{tab4} (corresponds to the right panel of Fig.\ref{b6}). From these two tables we can see that the biggest error is roughly $2.64\times10^{-2}$ in the Table \ref{tab4} for $h=0.8$ and $l_1=3$. This error is still very small, therefore, we can confirm that the two methods (machine learning with neural networks and diagonalization of Hamiltonian with SVD) match each other within errors. We attribute this $h$-dependent error to the underfitting phenomenon in the RBM approach. We think that as the system approaches criticality, the emergence of long-range correlations makes the wavefunction representation increasingly challenging, leading to the underfitting. This issue could potentially be mitigated by increasing the network width. Furthermore, from Fig.\ref{interval} and Fig.\ref{b6}, we observe that the multi-interval entanglement entropy exhibits greater sensitivity to this error compared to the bipartite entanglement entropy. On the other hand, it is noted that this kind of $h$-dependent error can also be found in other literatures, such as in the Fig. 3A of Ref. \cite{carleo2017solving}.

\section{Conclusions and Discussions}
\label{conclusion}
We studied the second order Renyi entropy $S_2$ for the one-dimensional TFQIM with two, three and four disjoint intervals from the improved swapping operations. In particular, we have used two different methods to compute $S_2$ from the expectation values of the swapping operator $S^{(2)}_{\rm wap}$. One method is the machine learning method with neural networks; The other one is the diagonalization of the Hamiltonian with SVD method. Consequently, we found that the results obtained from these two methods matched each other very well within errors, indicating that the machine learning methods with neural networks can be applied well in computing the Renyi entropy. By varying the magnetic field strength from smaller values to larger values, the phase of the spin chain will change from ferromagnetic to paramagnetic phase. From the results of Renyi entropy, we observed that as magnetic field was small, the value of $S_2$ was close to the constant $\ln2$. This behavior reflects the emergence of two degenerate ground state configurations with all spins point up or down in the thermodynamic limit, although in our case the ground state is unique due to the finite-size effect. As the magnetic field increased, the Renyi entropy increased as well until the magnetic field arrived the critical point $h_c=1$. However, as magnetic field exceeds the critical point, the phase of the system enters the paramagnetic phase. Therefore, the Renyi entropy will decrease as magnetic field goes beyond the critical value. As expected, if the magnetic field goes to infinity, the entropy will vanish since there will be only one configuration of the system. That is all the spins will point to the $x$-direction. 

The first significance of our paper is that we can not only study the Renyi entropy with two disjoint intervals, but also we can investigate the Renyi entropy with more disjoint intervals, such as three and four intervals. The later cases are seldom studied in the existing literatures. The second significance of our paper is that we have used the state-of-the-art machine learning methods with neural networks to study the Renyi entropy with more disjoint intervals. The results obtained from the machine learning methods are consistent with those from the SVD method. We also found that in the case of Fig.\ref{interval} with two disjoint intervals, the two methods match each other with high precision. However, in the case of Fig.\ref{b6} with three and four disjoint intervals, the errors between the two methods become larger although the errors are still acceptable.  We speculate that if there are too many disjoint intervals, the sampling approximations of the quantum states of the system from the machine learning methods may become not too accurate. The third significance of our paper is that our two methods can not only study the Renyi entropy at the critical point which is in conformal field theory, but also they can study the Renyi entropy beyond the critical regime. These two methods provide new avenues to study the Renyi entropy with multiple disjoint intervals.  

Meanwhile, it should also be noted that the swapping operation for computing Renyi entropy is viable for various states, not limited to the ground states. For instance, it may apply in the excited states or in the dynamical states. In our paper we only focus on the second order Renyi entropy $S_2$ in the one-dimensional TFQIM model, however, in principle it can be extended to higher order Renyi entropy $S_m$ in higher dimensions. We leave them for future work.

\section*{Acknowledgements}
 This work was partially supported by the National Natural Science Foundation of China (Grants No.12175008).

\appendix
\setcounter{equation}{0}
\setcounter{figure}{0}
\setcounter{table}{0}
\renewcommand{\theequation}{S.\arabic{equation}}
\renewcommand{\thefigure}{S.\arabic{figure}}


\section{Computational details of swapping operation}
\label{appendixa}

In this Appendix, we will show the computational details of arbitrary $m$-th order Renyi entropy $S_m$ with $n$ disjoint intervals. $S_m$ is defined as,
\begin{equation}
	S_{m}=\frac{1}{1-m}\ln\text{Tr}(\rho_A^m).
\end{equation}
The subsystem $A$ is $A=\cup_i a_i$ while the complement part $B=\cup_i b_i$ $(i=1, 2,\cdots n)$. Their arrangements are like those in the preceding section, i.e., $a_1b_1a_2b_2\cdots a_nb_n$. Under this partition, the wave function of the entire system can be represented as
\begin{equation}
	|\Psi\rangle=\sum_{\textbf{ab}}C_{a_1b_1\cdots a_nb_n}\otimes_{i=1}^n(|a_i\rangle|b_i\rangle),
\end{equation}	
in which $\sum_{\textbf{ab}}$ means $\sum_{\substack{a_1a_2\cdots a_n\\b_1b_2\cdots b_n}}$ and $\otimes_{i=1}^n$ is the $n$ times of the tensor product of the states $|a_i\rangle|b_i\rangle$. 
To obtain the value of $\text{Tr}(\rho_A^m)$, one can construct the $m$-th order swapping operator ${S}_{\rm wap}^{(m)}$ acting the $m$ copies of the states of the system. The rule is like,
\begin{eqnarray}
	&&{S}^{(m)}_{\rm wap}\otimes_{j=1}^m|\Psi^j\rangle={S}^{(m)}_{\rm wap}\otimes_{j=1}^m\left[\sum_{\textbf{a}^j\textbf{b}^j}C_{a_1^jb_1^j\cdots a_n^jb_n^j}\otimes_{i=1}^n(|a_i^j\rangle|b_i^j\rangle)\right]\nonumber\\
	&=&\otimes_{j=1}^{m-1}\left[\sum_{\textbf{a}^j\textbf{b}^j}C_{a_1^jb_1^j\cdots a_n^jb_n^j}\otimes_{i=1}^n(|a_i^{j+1}\rangle|b_i^j\rangle)\right]\otimes\left[\sum_{\textbf{a}^m\textbf{b}^m}C_{a_1^mb_1^m\cdots a_n^mb_n^m}\otimes_{k=1}^n(|a_{k}^1\rangle|b_{k}^m\rangle)\right].
\end{eqnarray}
In the above procedure, the swapping operator ${S}_{\rm wap}^{(m)}$ has the effect of replacing the state $|a_i^{j+1}\rangle$ in the subsystem $A^{j+1}$ in the $(j+1)$-th replica, with the state $|a_i^{j}\rangle$ in the subsystem $A^j$ in the $j$-th replica. This replacement is done for every two adjacent replicas. It should be noted that the subsystem $A^1$ of the first replica is replaced with the subsystem $A^m$ of the last replica. This procedure is similar to the replica trick in quantum field theory as we have discussed in the main text.

Under this process, the expectation value of the swapping operator can be calculated as,
\begin{eqnarray}\label{smop}
	\langle S_{\rm wap}^{(m)}\rangle&=&\langle\otimes_{j'=1}^m\Psi^{j'}|S_{\rm wap}^{(m)}|\otimes_{j=1}^m\Psi^{j}\rangle\nonumber\\
	&=&\otimes_{j'=1}^m[\sum_{\textbf{a}'^{j'}\textbf{b}'^{j'}}C^*_{{a'}_1^{j'}{b'}_1^{j'}\cdots {a'}_n^{j'}{b'}_n^{j'}}\otimes_{i'=1}^n(\langle {a'}_{i'}^{j'}|\langle {b'}_{i'}^{j'}|)]\otimes_{j=1}^{m-1}[\sum_{\textbf{a}^j\textbf{b}^j}C_{a_1^jb_1^j\cdots a_n^jb_n^j}\otimes_{i=1}^n(|a_i^{j+1}\rangle|b_i^j\rangle)]\nonumber\\
	&&\otimes[\sum_{\textbf{a}^m\textbf{b}^m}C_{a_1^mb_1^m\cdots a_n^mb_n^m}\otimes_{k=1}^n(|a_k^1\rangle|b_k^m\rangle)]\nonumber\\
	&=&\prod_{j,j'=1}^{m-1}\sum_{\substack{\textbf{a}^j\textbf{b}^j\\\textbf{a}'^{j'}\textbf{b}'^{j'}}}[C^*_{{a'}_1^{j'}{b'}_1^{j'}\cdots {a'}_n^{j'}{b'}_n^{j'}}C_{a_1^jb_1^j\cdots a_n^jb_n^j}\prod_{i=1}^n(\delta_{{a'}_i^{j'}a_i^{j+1}}\delta_{{b'}_i^{j'}b_i^j})]\nonumber\\
	&&\times\sum_{\substack{\textbf{a}^m\textbf{b}^m\\\textbf{a}'^m\textbf{b}'^m}}[C^*_{{a'}_1^m{b'}_1^m\cdots{a'}_n^m{b'}_n^m}C_{a_1^mb_1^m\cdots a_n^mb_n^m}\prod_{k=1}^n(\delta_{{a'}_k^ma_k^1}\delta_{{b'}_k^mb_k^m})]\nonumber\\
	&=&\prod_{j=1}^{m-1}\sum_{\textbf{a}^j\textbf{b}^j}[C^*_{a_1^{j+1}b_1^j\cdots a_n^{j+1}b_n^j}C_{a_1^jb_1^j\cdots a_n^jb_n^j}]\times\sum_{\textbf{a}^m\textbf{b}^m}[C^*_{a_1^1b_1^m\cdots a_n^1b_n^m}C_{a_1^mb_1^m\cdots a_n^mb_n^m}].
\end{eqnarray}
In order to verify $\text{Tr}(\rho_A^m)$ is equal to the expectation value of ${S}_{\rm wap}^{(m)}$, we need to know the density matrix $\rho$ of the whole system,
\begin{equation}
	\rho=|\Psi\rangle\langle\Psi|=\sum_{\textbf{abcd}}C_{a_1b_1\cdots a_nb_n}C^*_{c_1d_1\cdots c_nd_n}\otimes_{i=1}^n(|a_i\rangle|b_i\rangle)\otimes_{i'=1}^n(\langle c_{i'}|\langle d_{i'}|).
\end{equation}
Then, the reduced density matrix $\rho_A$ for subsystem A is obtained by tracing out the subsystem $B$,
\begin{eqnarray}
	\rho_A&=&\text{Tr}_B(\rho)=\sum_{\textbf{b}'}\otimes_{j=1}^n(\langle b'_j|)\rho\otimes_{j'=1}^n(|b'_{j'}\rangle)\nonumber\\
	&=&\sum_{\textbf{b}'}\sum_{\textbf{abcd}}C_{a_1b_1\cdots a_nb_n}C^*_{c_1d_1\cdots c_nd_n}\otimes_{j=1}^n(\langle b'_j|)\otimes_{i=1}^n(|a_i\rangle|b_i\rangle)\otimes_{i'=1}^n(\langle c_{i'}|\langle d_{i'}|)\otimes_{j'=1}^n(|b'_{j'}\rangle)\nonumber\\
	&=&\sum_{\textbf{abcdb}'}C_{a_1b_1\cdots a_nb_n}C^*_{c_1d_1\cdots c_nd_n}\prod_{j=1}^n(\delta_{b'_jb_j})\prod_{j'=1}^n(\delta_{d_{j'}b'_{j'}})\otimes_{i=1}^n(|a_i\rangle)\otimes_{i'=1}^n(\langle c_{i'}|)\nonumber\\
	&=&\sum_{\textbf{abc}}C_{a_1b_1\cdots a_nb_n}C^*_{c_1b_1\cdots c_nb_n}\otimes_{i=1}^n(|a_i\rangle)\otimes_{i'=1}^n(\langle c_{i'}|).
\end{eqnarray}
Next, we need to calculate the $m$-th power of reduced density matrix $\rho_A$,
\begin{eqnarray}
	\rho_A^m&=&\prod_{j=1}^m\left[\sum_{\textbf{a}^j\textbf{b}^j\textbf{c}^j}C_{a_1^jb_1^j\cdots a_n^jb_n^j}C^*_{c_1^jb_1^j\cdots c_n^jb_n^j}\otimes_{i=1}^n(|a_i^j\rangle)\otimes_{i'=1}^n(\langle c_{i'}^j|)\right]\nonumber\\
	&=&\prod_{j=1}^{m-1}\left[\sum_{\textbf{a}^j\textbf{b}^j\textbf{c}^j}C_{a_1^jb_1^j\cdots a_n^jb_n^j}C^*_{c_1^jb_1^j\cdots c_n^jb_n^j}\prod_{i=1}^n(\delta_{c_i^ja_i^{j+1}})\right]\nonumber\\
	&&\times\sum_{\textbf{a}^m\textbf{b}^m\textbf{c}^m}\left[C_{a_1^mb_1^m\cdots a_n^mb_n^m}C^*_{c_1^mb_1^m\cdots c_n^mb_n^m}\otimes_{k=1}^n(|a_k^1\rangle)\otimes_{k'=1}^n(\langle c_{k'}^m|)\right]\nonumber\\
	&=&\prod_{j=1}^{m-1}\left[\sum_{\textbf{a}^j\textbf{b}^j}C_{a_1^jb_1^j\cdots a_n^jb_n^j}C^*_{a_1^{j+1}b_1^j\cdots a_n^{j+1}b_n^j}\right]\nonumber\\
	&&\times\sum_{\textbf{a}^m\textbf{b}^m\textbf{c}^m}\left[C_{a_1^mb_1^m\cdots a_n^mb_n^m}C^*_{c_1^mb_1^m\cdots c_n^mb_n^m}\otimes_{k=1}^n(|a_k^1\rangle)\otimes_{k'=1}^n(\langle c_{k'}^m|)\right].
\end{eqnarray}
Then, the trace of $\rho_A^m$ becomes,
\begin{eqnarray}
	\text{Tr}(\rho_A^m)&=&\prod_{j=1}^{m-1}[\sum_{\textbf{a}^j\textbf{b}^j}C_{a_1^jb_1^j\cdots a_n^jb_n^j}C^*_{a_1^{j+1}b_1^j\cdots a_n^{j+1}b_n^j}]\times\!\!\!\!\!\sum_{\textbf{a}^m\textbf{b}^m\textbf{c}^m}\!\!\![C_{a_1^mb_1^m\cdots a_n^mb_n^m}C^*_{c_1^mb_1^m\cdots c_n^mb_n^m}\prod_{k=1}^n(\delta_{a_k^1c_k^m})]\nonumber\\
	&=&\prod_{j=1}^{m-1}[\sum_{\textbf{a}^j\textbf{b}^j}C_{a_1^jb_1^j\cdots a_n^jb_n^j}C^*_{a_1^{j+1}b_1^j\cdots a_n^{j+1}b_n^j}]\times\!\!\!\!\sum_{\textbf{a}^m\textbf{b}^m}\!\![C_{a_1^mb_1^m\cdots a_n^mb_n^m}C^*_{a_1^1b_1^m\cdots a_n^1b_n^m}]=\langle S_{\rm wap}^{(m)}\rangle.
\end{eqnarray}
From the Eq.\eqref{smop} we see that the trace $\text{Tr}(\rho_A^m)$ is exact equal to $\langle S_{\rm wap}^{(m)}\rangle$. 
Therefore, the computation of the $m$-th Renyi entropy $S_m$ with $n$ disjoint intervals is transformed to computing the expectation values of the swapping operator,  
\begin{equation}
	S_{m}=\frac{1}{1-m}\ln(\langle S_{\rm wap}^{(m)}\rangle).
\end{equation}

\section{Machine learning methods with neural networks}
\label{appendixb}
\subsection{Energy minimization with stochastic reconfiguration}
For a given Hamiltonian $H$, the expectation vale of energy can be estimated with neural network quantum state as
\begin{equation}
E(\mathcal{W})=\frac{\langle \Psi_{NN}(\mathcal{W})|H|\Psi_{NN}(\mathcal{W})\rangle}{\langle \Psi_{NN}(\mathcal{W})|\Psi_{NN}(\mathcal{W})\rangle}.
\end{equation}
Since the ground state has the lowest energy, for approximating the ground state, we can minimize the energy by optimizing $\mathcal{W}$ with the well-known stochastic reconfiguration (SR) method which has been studied extensively \cite{PhysRevLett.80.4558,PhysRevB.64.024512,10.1063/1.2437215,Becca_Sorella_2017,carleo2017solving}.
As we change the parameters slightly with $\mathcal{W}_k\to\mathcal{W}_k'=\mathcal{W}_k+\delta\mathcal{W}_k, k=1,2,3\dots p$, the new state can be expanded as
\begin{equation}
	|\Psi'(\mathcal{W}+\delta\mathcal{W})\rangle=\delta\mathcal{W}_0|\Psi(\mathcal{W})\rangle+\sum_{k'=1}^{p}\delta\mathcal{W}_{k'}\frac{\partial}{\partial\mathcal{W}_{k'}}|\Psi(\mathcal{W})\rangle.
\end{equation}
In order to estimate the energy with configuration samples later, we can rewrite $|\Psi\rangle$ in the configuration representation, in which the configurations $|s\rangle$ are complete bases of the Hilbert space,
\begin{eqnarray}
	|\Psi'(\mathcal{W'})\rangle&&=\delta\mathcal{W}_0|\Psi(\mathcal{W})\rangle+\sum_{k'=1}^{p}\delta\mathcal{W}_{k'}\frac{\partial}{\partial\mathcal{W}_{k'}}\sum_s|s\rangle\langle s|\Psi(\mathcal{W})\rangle\\
	&&=\delta\mathcal{W}_0|\Psi(\mathcal{W})\rangle+\sum_{k'=1}^{p}\delta\mathcal{W}_{k'}\sum_s\frac{\partial\langle s|\Psi(\mathcal{W})\rangle}{\partial\mathcal{W}_{k'}}|s\rangle\\
	&&=\delta\mathcal{W}_0|\Psi(\mathcal{W})\rangle+\sum_{k'=1}^{p}\delta\mathcal{W}_{k'}\sum_s\frac{\partial\ln\langle s|\Psi(\mathcal{W})\rangle}{\partial\mathcal{W}_{k'}}|s\rangle\langle s|\Psi(\mathcal{W})\rangle.
\end{eqnarray}
The variational derivatives with respect to the parameter $\mathcal{W}_k$ can be defined as
\begin{equation}
	\hat{\mathcal{O}}_k=\left\{
	\begin{aligned}
		&{\bf1},~~~~~~~~~~~~~~~~~~~~~~~\! {\rm for} ~k=0\\
		&\Sigma_s\tfrac{\partial\ln\langle s|\Psi(\mathcal{W})\rangle}{\partial\mathcal{W}_k}|s\rangle\langle s|,~~{\rm for} ~k\neq 0
	\end{aligned}
	\right.
\end{equation}
Therefore, the new state can be rewritten as operator $\hat{\mathcal{O}}_k$ acting on the original state,
\begin{equation}
	|\Psi'(\mathcal{W}')\rangle=\sum_{k'=0}^{p}\delta\mathcal{W}_{k'}\mathcal{O}_{k'}|\Psi(\mathcal{W})\rangle.
\end{equation}
Next, we need to match the operator $\hat{\mathcal{O}}_k$ to a step of iterations which can decrease the energy of state. We can choose the power method introduced in \cite{ruger2013implementation} to act the operator $(\Lambda-H)$ repeatedly to the state, where $\Lambda$ is a large enough number,
\begin{eqnarray}
	&|\Psi(\mathcal{W})\rangle=\sum\limits_{n=0}^{\infty}\langle n|\Psi(\mathcal{W})\rangle|n\rangle,\\
	&(\Lambda-H)^N|\Psi(\mathcal{W})\rangle=\sum\limits_{n=0}^{\infty}(\Lambda-E_n)^N\langle n|\Psi(\mathcal{W})\rangle|n\rangle.
\end{eqnarray}
As $\Lambda > E_n$ for arbitrary $n$, the ground state will become dominant with iterations, since the energy of ground state $E_0$ is the minimum eigenenergy. 
In one iteration, we should represent the new state which is the state from the last iteration acted by the operator $(\Lambda-H)$. Therefore, we can change the parameters $\mathcal{W}$ to match the changed state $|\Psi'(\mathcal{W}')\rangle$ and the new state, 
\begin{equation}
	|\Psi'(\mathcal{W}')\rangle=(\Lambda-H)|\Psi(\mathcal{W})\rangle=\sum_{k'=0}^{p}\delta\mathcal{W}_{k'}\hat{\mathcal{O}}_{k'}|\Psi(\mathcal{W})\rangle.
\end{equation}
Projecting the new state onto $\langle\Psi(\mathcal{W})|$ and $\langle\Psi(\mathcal{W})|\hat{\mathcal{O}}_k^\dagger$, we reach
\begin{eqnarray}
	&\langle\Psi(\mathcal{W})|\hat{\mathcal{O}}_{k}^{\dagger}(\Lambda-H)|\Psi(\mathcal{W})\rangle=\sum\limits_{k'=0}^p\langle\Psi(\mathcal{W})|\hat{\mathcal{O}}_{k}^{\dagger}\delta\mathcal{W}_{k'}\hat{\mathcal{O}}_{k'}|\Psi(\mathcal{W})\rangle, \\
	&\langle\Psi(\mathcal{W})|(\Lambda-H)|\Psi(\mathcal{W})\rangle=\sum\limits_{k'=0}^{p}\langle\Psi(\mathcal{W})|\delta\mathcal{W}_{k'}\hat{\mathcal{O}}_{k'}|\Psi(\mathcal{W})\rangle,
\end{eqnarray}
in which the left-hand sides can be rewritten as
\begin{eqnarray}
	&\Lambda\langle\hat{\mathcal{O}}_k^\dagger\rangle-\langle\hat{\mathcal{O}}_k^\dagger H\rangle=\langle\hat{\mathcal{O}}_k^\dagger\rangle\delta\mathcal{W}_0+\sum\limits_{k'=1}^{p}\langle\hat{\mathcal{O}}_k^\dagger\hat{\mathcal{O}}_{k'}\rangle\delta\mathcal{W}_{k'}\label{eqb},\\
	&\Lambda-\langle H\rangle=\delta\mathcal{W}_0+\sum\limits_{k'=1}^{p}\langle\hat{\mathcal{O}}_{k'}\rangle\delta\mathcal{W}_{k'}\label{eqa},
\end{eqnarray}
where $\langle\cdots\rangle$ represents the expect value in the configuration $s$. From Eq.\eqref{eqa}, we can obtain $\delta\mathcal{W}_0$ as
\begin{equation}
	\delta\mathcal{W}_0=\Lambda-\langle H\rangle-\sum\limits_{k'=1}^{p}\langle\hat{\mathcal{O}}_{k'}\rangle\delta\mathcal{W}_{k'}.
\end{equation}
And substitute it into Eq.\eqref{eqb}, we arrive at
\begin{eqnarray}
	\Lambda\langle\hat{\mathcal{O}}_k^\dagger\rangle-\langle\hat{\mathcal{O}}_k^\dagger H\rangle&=&\langle\hat{\mathcal{O}}_k^\dagger\rangle(\Lambda-\langle H\rangle-\sum\limits_{k'=1}^{p}\langle\hat{\mathcal{O}}_{k'}\rangle\delta\mathcal{W}_{k'})+\sum\limits_{k'=1}^{p}\langle\hat{\mathcal{O}}_k^\dagger\hat{\mathcal{O}}_{k'}\rangle\delta\mathcal{W}_{k'},\\
	\Longrightarrow \langle\hat{\mathcal{O}}_k^\dagger\rangle\langle H\rangle-\langle\hat{\mathcal{O}}_k^\dagger H\rangle&=&\sum\limits_{k'=1}^{p}(\langle\hat{\mathcal{O}}_k^\dagger\hat{\mathcal{O}}_{k'}\rangle-\langle\hat{\mathcal{O}}_k^\dagger\rangle\langle\hat{\mathcal{O}}_{k'}\rangle)\delta\mathcal{W}_{k'}.
\end{eqnarray}
For convenience, we can define the generalized force $f_k$ and covariance matrix $s_{kk'}$ as
\begin{eqnarray}
	f_k&\equiv&\langle\hat{\mathcal{O}}_k^\dagger H\rangle-\langle\hat{\mathcal{O}}_k^\dagger\rangle\langle H\rangle,\\
	s_{kk'}&\equiv&\langle\hat{\mathcal{O}}_k^\dagger\hat{\mathcal{O}}_{k'}\rangle-\langle\hat{\mathcal{O}}_k^\dagger\rangle\langle\hat{\mathcal{O}}_{k'}\rangle.
\end{eqnarray}
Therefore, the change of the parameters can be obtained as
\begin{equation}
	\delta\mathcal{W}_{k'}=\sum_k-(s^{-1})_{k'k}f_k.
\end{equation}
Now, we will show how to calculate the expectation values for neural network quantum state in the main text. First, the state can be described as
\begin{equation}
	\Psi(s,\mathcal{W})\equiv\langle s|\Psi(\mathcal{W})\rangle=\exp\left[\sum_ja_js_j\right]\prod_i2\cosh\left[b_i+\sum_jw_{ij}s_j\right].
\end{equation}
The values of $\langle\hat{\mathcal{O}}_k\rangle$ and $\langle H\rangle$ can be calculated by summing up the local values of $\mathcal{O}_k(s)$ and $E_{loc}(s)$ for all configurations according to its weight $|\Psi(s)|^2/\langle\Psi|\Psi\rangle$, where $\mathcal{O}_k(s)$ and $E_{loc}(s)$ are defined as,
\begin{eqnarray}
	\mathcal{O}_{a_j}(s)\equiv&&\tfrac{\partial\ln\Psi(s,\mathcal{W})}{\partial\mathcal{W}_{a_j}}=s_j,\\
	\mathcal{O}_{b_i}(s)\equiv&&\tfrac{\partial\ln\Psi(s,\mathcal{W})}{\partial\mathcal{W}_{b_i}}=\tanh[b_i+\sum_jw_{ij}s_j],\\
	\mathcal{O}_{w_{ij}}(s)\equiv&&\tfrac{\partial\ln\Psi(s,\mathcal{W})}{\partial\mathcal{W}_{w_{ij}}}=s_j\tanh[b_i+\sum_jw_{ij}s_j],\\
	E_{loc}(s)\equiv&&\tfrac{\langle s|H|\Psi(s,\mathcal{W})\rangle}{\Psi(s,\mathcal{W})}.
\end{eqnarray}

\subsection{Metropolis Hastings algorithm}
Since the number of configurations of the state is too large, we cannot calculate the expectation values accurately, instead we can only evaluate it through samples. Metropolis Hastings algorithm is an importance sampling method that can give an unbiased estimation of the expectation values \cite{chib1995understanding}.
 In the beginning, we generate a random configuration. Then a series of configurations are produced from this random configuration by the rules that the new configuration is chosen by flipping one site in the old configuration randomly, and accepting it by the probability,  
\begin{equation}
	P_{\rm accept}(s^i\to s^{i+1})=\min\left(1,\big\lvert\tfrac{\Psi(s^{i+1})}{\Psi(s^i)}\big\rvert^2\right)
\end{equation}
We can produce a Markovian chain of configurations by iterating the above operations. When the chain is long enough, the last configuration will not be affected by the initial choice, then we can add it to the sample. The frequency of configuration in the sample set obtained in this method will follow the probability $|\Psi(s)|^2/\langle\Psi|\Psi\rangle$ from the quantum state.
\\

\normalem
\bibliographystyle{JHEP}
\bibliography{ref1.bib}

\end{document}